\documentstyle[11pt,epsf]{article}
\newtheorem{theorem}{Theorem}

\newcommand {\dfn} {\stackrel{\Delta} {=}}

\newcommand {\eas} {\stackrel{\mbox{a.s.}} {=}}

\newcommand {\bp} {\mbox{\boldmath $p$}}

\newcommand {\br} {\mbox{\boldmath $r$}}
\newcommand {\bs} {\mbox{\boldmath $s$}}
\newcommand {\bt} {\mbox{\boldmath $t$}}
\newcommand {\bu} {\mbox{\boldmath $u$}}
\newcommand {\bv} {\mbox{\boldmath $v$}}

\newcommand {\bx} {\mbox{\boldmath $x$}}
\newcommand {\by} {\mbox{\boldmath $y$}}
\newcommand {\bz} {\mbox{\boldmath $z$}}

\newcommand {\bE} {\mbox{\boldmath $E$}}

\newcommand {\bS} {\mbox{\boldmath $S$}}
\newcommand {\bT} {\mbox{\boldmath $T$}}
\newcommand {\bU} {\mbox{\boldmath $U$}}

\newcommand {\bX} {\mbox{\boldmath $X$}}
\newcommand {\bY} {\mbox{\boldmath $Y$}}
\newcommand {\bZ} {\mbox{\boldmath $Z$}}
\newcommand{\calA}{{\cal A}}

\newcommand{\calE}{{\cal E}}

\newcommand{\calS}{{\cal S}}
\newcommand{\calT}{{\cal T}}

\newcommand{\calX}{{\cal X}}
\newcommand{\calY}{{\cal Y}}
\newcommand{\calZ}{{\cal Z}}
\begin{document}
\thispagestyle{empty}
\title{Joint Source--Channel Coding via Statistical Mechanics:
Thermal Equilibrium Between the Source and the Channel
\thanks{Part of this work was carried out during a visit in Hewlett--Packard
Laboratories, Palo Alto, CA, U.S.A., in the Summer of 2008.}
}
\author{Neri Merhav
}
\date{}
\maketitle

\begin{center}
Department of Electrical Engineering \\
Technion - Israel Institute of Technology \\
Haifa 32000, ISRAEL \\
\end{center}
\vspace{1.5\baselineskip}
\setlength{\baselineskip}{1.5\baselineskip}

\begin{abstract}
We examine the classical joint source--channel coding problem from the
viewpoint of statistical physics and demonstrate that 
in the random coding regime, the posterior
probability distribution of the source given the channel output
is dominated by source sequences, which exhibit a behavior that is highly parallel
to that of thermal equilibrium between two systems of particles that exchange
energy, where one system corresponds to the source and the other corresponds to the
channel. The thermodynamical entopies of the dual physical
problem are analogous to conditional and unconditional Shannon
entropies of the source, 
and so, their balance in thermal equilibrium yields a simple formula for the mutual information 
between the source and the channel output, that is induced by the
typical code in an ensemble of joint source--channel codes under certain
conditions. We also demonstrate how our results can be used in applications,
like the wiretap channel, and how can it be extended to multiuser scenarios,
like that of the multiple access channel.

\vspace{0.25cm}

\end{abstract}

\section{Introduction}

Consider the following two seemingly unrelated problems, which
serve as simple special cases of a more general setting we study later in this paper:

The first is an elementrary problem in statistical
physics: We have two subsystems of particles which are
brought into thermal equilibirium with each other as well
as with the environment (a heat bath) at temperature $T$.
The first subsystem consists of $N$
particles having magnetic moments (spins), $\{s_i\}$, each of which may
be oriented either in the direction 
of an applied external magnetic field $B$,
in which case $s_i=+1$, or in the opposite direction, in which case
$s_i=-1$, and its energy in both cases is given by $-s_iB$ (up to a certain 
multiplicative constant, which carries the appropriate physical units,
and which is irrelevant for the purpose of this discussion).
In the second subsystem, there are $n$ non--interacting particles
$\{s_i'\}_{i=1}^n$,
each one of which may lie in one of two possible states: the state $s_i'=0$,
in which the particle has zero energy, and the state $s_i'=1$, in which it
has energy $e_0$. 
What is the average energy possessed by each one of these subsystems
in equilibrium, as functions of $e_0$, $T$, $n$, $N$, and $B$?

The second problem is in Information Theory, in particular, it is in
joint source--channel coding, where some of
the notation used is deliberately chosen to be the same as
in the previous paragraph: A binary memoryless source 
generates a vector $\bs$ of symbols $(s_1,s_2,\ldots,s_N)$, 
$s_i\in\{+1,-1\}$, 
$i=1,\ldots,N$, with probabilities $q=\mbox{Pr}\{S_i=+1\}$
and $1-q=\mbox{Pr}\{S_i=-1\}$.
This vector is encoded into a binary channel codeword $\bx(\bs)$
of length $n$ and transmitted over a binary symmetric channel (BSC) with
a crossover probability $p < 1/2$, and a binary $n$--vector $\by$ is
received at the channel output. Consider the posterior distribution
$$P(\bs|\by)=\frac{P(\bs)W(\by|\bx(\bs))}{\sum_{\bs'}
P(\bs')W(\by|\bx(\bs'))}$$
where $P(\bs)$ and $W(\by|\bx)$ are the probability distributions that
govern the source and the channel, respectively, as described above.
Thus, clearly, $P(\bs|\by)$ is proportional to $P(\bs)W(\by|\bx(\bs))$, or
equivalently,
$\ln P(\bs|\by)$ is (within a term 
that is independent of $\bs$) given by $\ln P(\bs)+\ln W(\by|\bx(\bs))$. 
For a typical code drawn uniformly at random from the ensemble of codes,
what are the relative contributions of the source
and the channel to this quantity, for those vectors $\bs$ that dominate
$P(\bs|\by)$ (i.e., those that capture the vast majority of the
posterior probability)?

It turns out, as we shall see in Section \ref{bkgd} below,
that the two problems have virtually identical answers (in a sense
that will be made clear and precise therein), provided that the
parameters $T$ and $B$ of the first problem are related to the parameters $p$
and $q$ of the second problem by
\begin{equation}
\label{itp1}
p = \frac{\exp\{-e_0/kT\}}{1+\exp\{-e_0/kT\}}
\end{equation}
and
\begin{equation}
\label{itp2}
q=\frac{\exp\{B/kT\}}{2\cosh(B/kT)},
\end{equation}
or equivalently,
\begin{equation}
\label{itp3}
e_0=kT\ln\frac{1-p}{p}
\end{equation}
and
\begin{equation}
\label{itp4}
B=\frac{kT}{2}\ln\frac{q}{1-q},
\end{equation}
where $k$ is Boltzmann's constant.

Thermal equilibrium between the two subsystems 
in the above described physical problem, dictates a
certain balance between their thermodynamical entropies
in order to arrive at the maximum total entropy (by the second
law of thermodynamics) for the total energy possessed by the
entire system at the given temperature $T$. As the thermodynamical entropy, in
its statistical--mechanical definition, is intimately related to the Shannon
entropy, it turns out that this equilibrium relation between the thermodynamical entropies
of the physical problem, 
gives rise to an analogous relation between Shannon entropies pertaining to
the joint source--channel coding problem in the random coding regime.
In particular, it relates the
entropy of the source to its conditional entropy given the channel output,
whose difference is exactly the mutual information between the source
and the channel output. 
The final outcome of this is a simple
formula for calculating the mutual information rate between the input and the
output of a coded system for the typical code in a given ensemble under
certain conditions. This calculation builds strongly on the random energy
model (REM) of spin glasses due to Derrida \cite{Derrida80a, Derrida80b,
Derrida81}
and its relation to the random code ensemble (RCE) as described in
\cite{MM08}.

Clearly, under the regime of reliable communication, the mutual
information rate between the source and the channel output coincides
with the entropy rate of the source, as the conditional entropy rate of the
source given the channel output vanishes. Thus, the problem of calculating
the mutual information under reliable communication conditions is easy and
in fact, not quite interesting. The same calculation, however,
when the conditions of reliable communication are not met, appears less
trivial. But what would be the motivation for such a calculation?

Here are just a few examples that motivate this: Consider a
user that, in addition to its desired signal, receives also a relatively
strong interfering
signal (codeword), which is intended to other users, and which comes
from a codebook whose rate exceeds the capacity of this crosstalk channel
between the interferer and our user, so that the user cannot fully decode
this interference. Nonetheless, our user would like to learn as much as
possible on the interfering signal for many possible reasons: For example,
the user would like to learn the interference signal
in order to identify where it originates from, or in order to estimate it
and subtract it (intereference cancellation). The mutual information
rate, call it $I$,  between the interference signal and the channel output then gives
some assessment concerning the quality of this estimation. For one thing,
$D(I)$, where $D(\cdot)$ is the distortion--rate function of the source, 
is a lower bound to the distortion in estimating this signal. Moreover, if the
channel is Gaussian, one can calculate the {\it exact} minimum mean square error
(MMSE) from the mutual information rate $I$ by taking its derivative w.r.t.\ the
signal--to--noise ratio (SNR) \cite{GSV05}. Another application comes from
scenarios where the above described receiver is a hositle party (an
eavesdropper), from which one
would like to conceal information as much as possible. The natural setup, in
this context, is that of the wiretap channel (cf.\ \cite{Wyner75} as well as
many follow--up papers), where excess channel noise beyond capacity is
harnessed as an effective key that secures data communication.
As we show in the sequel, the mutual information rate
between the transmitted message and the eavesdropper, which suffers from this
excess noise, is strongly related to the equivocation, which is a customary
measure of security in Shannon--theoretic secrecy systems.

The outline of this paper is as follows. In Section 2, we establish notation
conventions. In Section 3, we provide some basic background of elementary
statistical physics, which will be needed in the sequel. In Section 4, we
derive our main result, which is a formula for the mutual information rate.
In Section 5, we demonstrate how it is applied for the wiretap channel, and
finally, in Section 6, we demonstrate how our results can be extended to
multiuser scenarios, like that of the multiple access channel.

\section{Notation Conventions}

Throughout this paper, scalar random 
variables (RV's) will be denoted by the capital
letters, like $S$, $X$, and $Y$, their sample values will be denoted by
the respective lower case letters, and their alphabets will be denoted
by the respective calligraphic letters.
A similar convention will apply to
random vectors and their sample values,
which will be denoted with same symbols with the bold face font.
Thus, for example, $\bX$ will denote a random $n$-vector $(X_1,\ldots,X_n)$,
and $\bx=(x_1,...,x_n)$ is a specific vector value in $\calX^n$,
the $n$-th Cartesian power of $\calX$. 
Sources and channels will be denoted generically by the letter $P$,$Q$,$M$ and
$W$. Whenever clarity and unambiguity will require it, these letters will be
subscripted by the names of the relevant RV's, following the standard
notation conventions in the literaure,
for example, $P_S$ will denote the probability
distribution of a random variable $S$, $P_{X|Y}$ will denote the conditional
probability distribution of $X$ given $Y$, and so on.
The cardinality of a finite set $\calA$ will be denoted by $|\calA|$.
Information theoretic quantities like entropies and mutual
informations will be denoted following the usual conventions
of the information theory literature.

\section{Background}
\label{bkgd}

In this section, we provide a brief account 
of the very basic background in statistical physics,
which is needed for this paper.

Consider a physical system with $N$ of particles,
which can be in a variety of microscopic states (`microstates'),
defined by combinations of of physical quantities associated with
these particles, e.g.,
positions, momenta, 
angular momenta, spins, etc., of all $N$ particles. 
For each such 
microstate of the system, which we shall
designate by a vector $\bs=(s_1,\ldots,s_N)$, there is an
associated energy, given by an {\it Hamiltonian} (energy function),
$\calE(\bs)$. For example, if $s_i=(\bp_i,\br_i)$, where
$\bp_i$ is the momentum vector of particle number $i$ and
$\br_i$ is its position vector, then classically, $\calE(\bs)=\sum_{i=1}^N
[\frac{\|\bp_i\|^2}{2m}+mgz_i]$, where $m$ is the mass of each particle,
$z_i$ is its height -- one of the
coordinates of $\br_i$, and
$g$ is the gravitation constant. 

One of the most
fundamental results in statistical physics (based on the 
law of energy conservation and
the basic postulate that all microstates of 
the same energy level are equiprobable)
is that when the system is in thermal 
equilibrium with its environment, the probability of a microstate $\bs$ is
given by the {\it Boltzmann--Gibbs} distribution
\begin{equation}
\label{bd}
P(\bs)=\frac{e^{-\beta\calE(\bs)}}{Z(\beta)}
\end{equation}
where $\beta=1/(kT)$, $k$ being Boltmann's contant and $T$ being temperature, and
$Z(\beta)$ is the normalization constant, 
called the {\it partition function}, which
is given by
$$Z(\beta)=\sum_{\bs} e^{-\beta\calE(\bs)}$$
or
$$Z(\beta)=\int d\bs e^{-\beta\calE(\bs)},$$
depending on whether $\bs$ is discrete or continuous. The role
of the partition function is by far deeper 
than just being a normalization factor, as
it is actually the key quantity from which many 
macroscopic physical quantities can be derived,
for example, the free energy\footnote{The free 
energy means the maximum work that the system
can carry out in any process of fixed temperature.
The maximum is obtained when the process is reversible (slow, quasi--static
changes in the system).}
is $-\frac{1}{\beta}\ln Z(\beta)$, the average internal
energy (i.e., the expectation of $\calE(\bs)$ where 
$\bs$ drawn is according (\ref{bd}))
is given by the negative derivative of $\ln Z(\beta)$, the heat capacity 
is obtained from
the second derivative, etc. One of the ways to 
obtain eq.\ (\ref{bd}), is as the maximum entropy
distribution under an energy constraint 
(owing to the second law of thermodynamics), where $\beta$ plays the role of
a Lagrange multiplier that controls this energy level.

Let us define the quantity:
\begin{equation}
\label{Omega}
\Omega_{N,\delta}(\epsilon)=\bigg|\left\{\bs:~
(\epsilon-\delta/2)N
\le \calE(\bs)\le 
(\epsilon+\delta/2)N\right\}\bigg|,
\end{equation}
and let us assume that the limit
$$\Sigma(\epsilon)=\lim_{\delta\to 0}
\lim_{N\to\infty}\frac{\ln\Omega_{N,\delta}(\epsilon)}{N}$$
exists and that $\Sigma(\epsilon)$ is a differentiable concave function.
$\Sigma(\epsilon)$ is the {\it entropy} of the physical system in its
statistical--mechanical definition. We will see shortly that it is
intimately related to the Shannon entropy associated with 
the Boltzmann--Gibbs probablity distribution $P(\bs)$
defined above.

To see why the concavity assumption makes sense,
note that at least when
$P(\bs)$ is a product distribution (namely, when $\calE(\bs)=
\sum_i\calE(s_i)$),
$$\Omega_{N_1+N_2,\delta}\left(\frac{N_1\epsilon_1+N_2\epsilon_2}{N_1+N_2}
\right)\geq \Omega_{N_1,\delta}(\epsilon_1)\cdot
\Omega_{N_2,\delta}(\epsilon_2)$$
since for every configuration $\bs$, where $N_1\le N$ particles have total
energy $N_1\epsilon_1$ and $N_2=N-N_1$ particles have total energy
$N_2\epsilon_2$, the total energy of all $N=N_1+N_2$ particles is
obviously $N_1\epsilon_1+N_2\epsilon_2$, but the converse is not true
since there are other ways to split the total energy of
$N_1\epsilon_1+N_2\epsilon_2$ between the two 
complementary subsets of particles.
Thus, taking the logarithm of both sides, 
dividing by $(N_1+N_2)$, then taking the limits of
$N_1,N_2\to\infty$ such that $N_1/N_2$ tends to a given
constant, and finally, taking the limit of $\delta\to 0$,
one readily observes that $\Sigma(\epsilon)$ is concave. An argument of the
same spirit can be exercised in somewhat more general situations, e.g.,
when $P(\bs)$ has a Markov structure (namely,
the physical system has some nearest--neighbor interactions), though
some more caution is required. 

Denoting 
$$\psi(\beta)=\lim_{N\to\infty}\frac{1}{N}
\ln\sum_{\bs}\exp\{-\beta\calE(\bs)\},$$
it is readily seen that
\begin{eqnarray}
\psi(\beta)&=&
\lim_{\delta\to 0}\lim_{N\to\infty}\frac{1}{N}\ln\left[\sum_{j\ge 0}
\Omega_{N,\delta}((j+1/2)\delta)\cdot\exp\{-N\beta j\delta]\}\right]\nonumber\\
&=&\sup_{\epsilon\ge 0}[\Sigma(\epsilon)-\beta\epsilon],
\end{eqnarray}
i.e., 
$\psi(\cdot)$ and $\Sigma(\cdot)$ are a Legendre--transform pair.
Since $\Sigma(\cdot)$ is assumed concave, then the inverse transform
relation
$$\Sigma(\epsilon)=\inf_{\beta\ge 0}[\beta\epsilon+\psi(\beta)],$$
holds true as well, and so the derivatives $\beta(\epsilon)\dfn
d\Sigma/d\epsilon$ and $\epsilon(\beta)=-d\psi/d\beta$
(which are the maximizer of $[\Sigma(\epsilon)-\beta\epsilon]$
and the minimizer of $[\beta\epsilon+\psi(\beta)]$, respectively),
are inverses of each other. It follows then that
$$\Sigma(\epsilon)=\psi(\beta)-\beta\cdot\frac{d\psi}{d\beta},$$
but as is readily seen, $-d \psi/d\beta$ is the average internal energy,
$\bE\{\calE(\bS)]\}$, where $\bE$ is
the expectation operator associated
with the Boltzmann distribution.
This, in turn, is readily verified to agree with the 
expression of the Shannon entropy rate $H(S)$
of the distribution $P(\bs)$,
\begin{eqnarray}
H(S)&=&\lim_{n\to\infty}\frac{1}{N}\bE
\left\{\ln\left[\frac{1}{P(\bS)}\right]\right\}\nonumber\\
&=&\lim_{n\to\infty}\frac{1}{n}\bE\left\{
\ln\left[\frac{Z(\beta)}{\exp\{-\beta\calE(\bS)\}}
\right]\right\}\nonumber\\
&=&\psi(\beta)+\beta\bE\{\calE(\bS)\}.
\end{eqnarray}
Thus, $\Sigma(\epsilon)=H(S)$ whenever $\beta$ and $\epsilon$ are
related by $\beta=\beta(\epsilon)$, or equivalently,
$\epsilon=\epsilon(\beta)$.
For a given $\beta$, the Boltzmann--Gibbs 
distribution has a sharp peak (for large $N$) at 
the level of $\epsilon(\beta)$. We then say that this value of $\epsilon$
is the {\it dominant} energy level: Not only is it the average energy, 
there is also a strong concentration of the probability about this value
as $N$ grows without bound.
The second law of thermodynamics asserts
that in an isolated system (which does not exchange energy with
its environment), the total entropy cannot decrease, and hence in
equilibrium, it reaches its maximum. 

Now, suppose that we have a physical system that is composed of two
subsystems, one having $N$ particles with microstates $\{\bs\}$
and Hamiltonian $\calE_1(\bs)$, and
the other has $n$ particles with microstates $\{\bs'\}$ 
and Hamiltonian $\calE_2(\bs')$. Let us
suppose that these two subsystems are in thermal contact 
and they both reside in a very large environment (heat bath) having
a fixed temperature $T=1/(k\beta)$.
The two subsystems are
allowed to exchange energy with each other as well as with the heat bath.
How is the total energy of the system split between the two subsystems?
An example of two such subsystems was described in the first few paragraphs
of the Introduction.

The partition function of the composite system is given by
$$Z(\beta)=\sum_{\bs,\bs'}\exp\{-\beta[\calE_1(\bs)+\calE_2(\bs')]\}$$
and so the dominant energy level, as we saw before, is the one that 
achieves the associated normalized log--partition function
$\psi(\beta)$, i.e., the solution $\epsilon_0$
to the equation $d\Sigma(\epsilon)/d\epsilon=\beta$, where $\Sigma(\epsilon)$
is the entropy of the combined system.
Let us confine attention now
to the set of combined 
microstates $\{(\bs,\bs')\}$ of the composite system which
have energy $(N+n)\epsilon_0$. More precisely,
assume that the ratio $n/N=\lambda$ is held fixed, 
so $(N+n)\epsilon_0=N(1+\lambda)\epsilon_0$, and let us define
$$\Omega_{N,n,\delta}(\epsilon_0)=\Bigg|\{(\bs,\bs'):~
N(1+\lambda)(\epsilon_0-\delta/2)\le \calE_1(\bs)+\calE_2(\bs')\le
N(1+\lambda)(\epsilon_0+\delta/2)\}\Bigg|.$$
Clearly, every configuration $(\bs,\bs')$ with energy
about $N(1+\lambda)\epsilon_0$ corresponds to some allocation of
of the energy in one subsystem and the remaining energy in the
other. Thus, defining $\Omega_{N,\delta}^{(1)}(\epsilon)$ and
$\Omega_{n,\delta}^{(2)}(\epsilon)$ as the enumerators 
of microstates with energy about $\epsilon$
in each one of the two subsystems individually (as defined in eq.\ 
(\ref{Omega})), we have, for $\hat{\delta}=\delta(1+\lambda)$:
$$\Omega_{N,n,\hat{\delta}}(\epsilon_0)=\sum_{j\ge 0}
\Omega_{N,\delta}^{(1)}((j+1/2)\delta)\Omega_{n,\delta}^{(2)}\left(
\frac{(1+\lambda)\epsilon_0-(j+1/2)\delta}{\lambda}\right).$$
Defining $\Sigma(\epsilon)$ as
$\lim_{\delta\to 0}\lim_{N\to\infty}[\ln
\Omega_{N,\lambda N,\hat{\delta}}(\epsilon)]/[N(1+\lambda)]$,
we find, after taking logarithms of both sides, dividing by
$N(1+\lambda)$, letting $N\to \infty$, and then $\delta\to 0$,
that $\Sigma(\epsilon_0)$ is given by the weighted supremal
convolution\footnote{The supremal convolution between two functions
$f(x)$ and $g(x)$ is generally defined as $h(x)=\sup_t[f(x-t)+g(t)]$. The
qualifier ``weighted'', in our context, refers to the fact that both
functions as well as their arguments are weighted by $1/(1+\lambda)$ and
$\lambda/(1+\lambda)$.}:
$$\Sigma(\epsilon_0)=\sup_{0\le\epsilon\le(1+\lambda)
\epsilon_0}\left[\frac{1}{1+\lambda}\cdot\Sigma_1(\epsilon)
+\frac{\lambda}{1+\lambda}\cdot
\Sigma_2\left(\frac{(1+\lambda)\epsilon_0-\epsilon}{\lambda}\right)\right].$$
Assuming that the maximum is achieved by
$\epsilon^*\in(0,(1+\lambda)\epsilon_0)$, it is characterized
by a vanishing derivative of the expression in the square brackets,
i.e., the solution to the equation
\begin{equation}
\label{thermaleq}
\Sigma_1'(\epsilon)=\Sigma_2'\left(\frac{(1+\lambda)\epsilon_0-
\epsilon}{\lambda}\right),
\end{equation}
where $\epsilon$ is the unknown, and
where $\Sigma_i'$ is the derivative of $\Sigma_i$, $i=1,2$. This equation
characterizes the thermal equilibrium between the two subsystems
and the heat bath. Now, the left--hand side is exactly $\beta$. Thus,
$\epsilon^*$, the per--particle energy share of the first subsystem
is the solution to the equation $\Sigma_1'(\epsilon)=\beta$
(or, equivalently, of eq.\ (\ref{thermaleq}), as said), and
the remaining energy per particle, 
$[(1+\lambda)\epsilon_0-\epsilon^*]/\lambda$ belongs to the other subsystem.

\vspace{0.25cm}

\noindent
{\it Comment.}
Returning to the example that opens the Introduction, a simple calculation
shows that the dominant energies are
$$H\cdot\bE\{\sum_{i=1}^NS_i\}=
NB\tanh\left(\frac{B}{kT}\right)$$
in the first subsystem, and 
$$e_0\cdot\bE\{\sum_{i=1}^nS_i'\}=
\frac{ne_0\exp\{-e_0/kT\}}{1+\exp\{-e_0/kT\}}$$
in the second subsystem. Thus,
$$\epsilon^*=B\tanh\left(\frac{B}{kT}\right)$$
and
$$\frac{(1+\lambda)\epsilon_0-\epsilon^*}{\lambda}=\frac{\exp\{-e_0/kT\}}{1+\exp\{-e_0/kT\}}.$$
In the parallel joint source--channel coding problem described in the
Introduction, and to be further studied in a more general setting
in the sequel, we have:
$\ln P(\bs)=(\frac{1}{2}\ln\frac{q}{1-q})\cdot\sum_{i=1}^Ns_i+\mbox{const}$,
and $\ln W(\by|\bx)=(\ln\frac{p}{1-p})\cdot\sum_{i=1}^n(x_i\oplus y_i)+
\mbox{const}$, with $\oplus$ denoting modulo 2 addition, the dominant
contribution to $P(\bs|\by)$ comes from those $\{\bs\}$ for which
$\sum_{i=1}^Ns_i$ is about its typical value $N[(+1)\cdot q+(-1)\cdot(1-q)]
=N(2q-1)=N\tanh(B/kT)$ (in analogy to the energy of the first subsystem above,
where we have used the relations (\ref{itp1})-(\ref{itp4}))
and $\sum_{i=1}^n(x_i\oplus y_i)$ is
about $np=n\exp\{-e_0/kT\}/[1+\exp\{-e_0/kT\}]$ (in analogy to
the energy of the second subsystem). 
Notice that these two typical contributions to
the log--posterior probability agree also with
the corresponding typical contributions, $\ln P(\bs_0)$ and
$\ln W(\by|\bx(\bs_0))$,
of the {\it real} message $\bs_0$
that was actually transmitted.
This is true regardless of whether the communication is reliable or not,
i.e., it continues to hold no matter whether the entropy rate of the
source is smaller or larger than $\lambda$ times the mutual information
between the input and the output of the channel.

Returning to the general discussion above, 
note that the same considerations continue to hold even if one
of the systems, say, the second one, has an effective {\it negative} entropy,
that is, $\Omega_{n,\delta}^{(2)}(
[(1+\lambda)\epsilon_0-\epsilon^*]/\lambda) < 1$,
which means that for each microstate $\bs$ of the first subsystem
with per--particle energy $\epsilon^*$, only {\it a fraction}
of the compatible combined microstates $\{(\bs,\bs')\}$ 
have noramilzed energy $\epsilon_0$. Of course,
$\Omega_{N,n,\hat{\delta}}(\epsilon^*)$ must be larger than 1.
In the sequel, we shall see that in the joint source--channel coding
problem, the source and the channel constitute a mechanism which is
highly parallel to that of equilibrium energy--sharing
between two subsystems in a heat bath, where the subsystem corresponding
to the channel has a negative effective thermodynamic entropy in this sense.

We should comment that in order to determine the energy sharing
between the two subsystems in the above discussion, it was not
necessary to consider how they thermally interact with each other and to
go through the weighted supremal convolution between their
entropies, as we did. We could have determined these energies simply
by considering the equilibrium of each one of the subsystems 
individually with the
heat bath,\footnote{When doing so, the other system then
becomes part of the heat bath anyway.} 
thus equating the derivative of each one of the
entropy functions to $\beta$. Nonetheless, we have deliberately chosen
to present the supremal convolution because in the sequel, it is this relation
that will lead to the derivation of the mutual information in the joint
source--channel coding problem.

\section{Formulation, Main Results and Discussion}

Consider an information source, $S_1,S_2,\ldots$, 
whose symbols $\{S_i\}$ take on values in a finite
alphabet $\calS$. The source is characterized
by a sequence of probability distributions, 
$P(\bs)$, $\bs\dfn (s_1,\ldots,s_N)$, where $N=1,2,\ldots$.
Consider next a discrete memoryless channel (DMC), which 
is characterized by
a matrix of single--letter transition probabilities
$\{W(y|x),~x\in\calX,~y\in\calY\}$, where $\calX$ and $\calY$ are
finite alphabets. The operation rate of the channel relative to
the source is $\lambda$ channel uses per source symbol, which means
that while the source produces an $N$--vector $\bs=(s_1,\ldots,s_N)\in\calS^N$, the
channel conveys $n$ channel symbols, namely, it receives an
$n$--vector $\bx=(x_1,\ldots,x_n)\in\calX^n$ and outputs an $n$--vector
$\by=(y_1,\ldots,y_n)\in\calY^n$, where $n=\lambda N$. The parameter $\lambda$
is referred to as the {\it bandwidth expansion factor} of the channel relative
to the source.

For the sake of convenience in drawing the
analogy with statistical mechanics,
we will think of both the source and the channel as
Boltzmann distributions with certain Hamiltonians at a certain
common inverse temperature $\beta$, that is, 
$P(\bs)$ is proportional to $\exp\{-\beta\calE_S(\bs)\}$ and
$W(y|x)$ is proportional to $\exp\{-\beta\calE_C(x,y)\}$, where
$\calE_S(\cdot)$ and $\calE_C(\cdot,\cdot)$ are the Hamiltonians of the source
and the channel, respectively. 
For a pair of $n$--vectors $\bx$ and $\by$, we will denote
$W(\by|\bx)=\prod_{i=1}^n W(y_i|x_i)$, and keep in mind that it is proportional to
$\exp\{-\beta\calE_C(\bx,\by)\}$, where
$\calE_C(\bx,\by)\dfn\sum_{i=1}^n\calE_C(x_i,y_i)$.
Clearly, there is no loss of generality in this
representation of the source and the channel 
since there is always at least one way of doing this: For
example, one can simply
take $\beta=1$, $\calE_S(\bs)=-\ln P(\bs)$, and $\calE_C(x,y)=-\ln W(y|x)$.
The point is, however, that by doing this we have slightly extended the scope:
instead of one source and one channel, we are actually considering a family of
sources and channels, both indexed by a common parameter $\beta$, that
controls the degree of uniformity or skewedness of the distribution.

An $(N,n)$ {\it joint source--channel code}, for the above
defined source and channel, is a mapping from
the set $\calS^N$
to $\calX^n$.
Every source string $\bs$ is mapped into a channel input vector
$\bx\dfn (x_1,\ldots,x_n)$, and when we wish
to emphasize the dependence of $\bx$ on $\bs$, we denote it
as $\bx(\bs)$. The code is assumed to be selected at random, where
for each $\bs$, the codeword $\bx(\bs)$ is drawn under
a distribution\footnote{A more general model would allow a distribution
$M$ that depends on $\bs$. For example, if $\calS^N$ can be naturally divided
into type classes (like in te case of memoryless sources, Markov sources,
etc.), then it is plausible to let $M$ depend on the type class of $\bs$.
However, among all sequences in $\calS^N$,
the important ones are those that are typical to the source
(others can be ignored in the
large $N$ limit),
which are equiprobable in the
exponential scale, and so, the distribution $M$ for all of them can be
taken to be the same without loss of asymptotic optimality.}
$M(\bx)$, independently\footnote{The independence assumption
is made here mostly for the sake of simplicity. It can be
somewhat relaxed
as long as the concentration properties specified below continue to hold.}
of all other codewords.
The receiver estimates $\bs$ by applying a certain
function on the received channel output sequence $\by\dfn(y_1,\ldots,y_n)$,
i.e., it implements a function from $\calY^n$ to $\calS^N$, which
will be denoted by $\hat{\bs}=\hat{\bs}(\by)$.
In some applications, the receiver (or the observer) may not
necessarily attempt at full--fledged decoding of the message, but
may opt to merely estimate a certain function of the source sequence
(e.g., some statistic such as its composition).

Our study of the mutual information induced by the joint
source--channel code will be strongly based on the posterior distribution,
which, for a given (randomly selected) code, is defined as:
\begin{eqnarray}
\label{posterior}
P_\beta(\bs|\by)&=&
\frac{P(\bs)W(\by|\bx(\bs))}{\sum_{\bs'\in\calS^N}P(\bs')W(\by|\bx(\bs'))}\nonumber\\
&=&\frac{\exp\{-\beta[\calE_S(\bs)+\calE_C(\bx(\bs),\by)]\}}
{\sum_{\bs'}\exp\{-\beta[\calE_S(\bs')+\calE_C(\bx(\bs'),\by)]\}}.
\end{eqnarray}
On a technical note, observe 
that since the posterior distribution is given by a ratio, this allows
slighlty more freedom in the definition of the Hamiltonians $\calE_S$ and
$\calE_C$, as certain common constants in the numerator and the denominator
may cancel each other. For example, if the source is binary and memoryless,
as described in the example given in the Introduction,
then $P(\bs)$ is proportional to
$\exp\{-(\frac{1}{2}\ln\frac{1-q}{q})\sum_{i=1}^Ns_i\}$, and so
one can define $\calE_S(\bs)$ to be proportional to
$\sum_{i=1}^Ns_i$, where the factor $\frac{1}{2}\ln\frac{1-q}{q}$ can
be split between a part that is absrobed 
in the Hamiltonian itself and a part that is attributed to the inverse temperature parameter
$\beta$. A similar comment applies to the channel, but here some more caution
is required since, in general, the constant of proportionality that relates
$W(\by|\bx)$ and $\exp\{-\beta\calE_C(\bx,\by)\}$ may depend on $\bx$, unless
the code is of constant composition and/or the channel is symmetric in the
sense that $\sum_y \exp\{-\beta\calE_C(x,y)\}$ is independent of $x$ for all
$\beta$ (which is the case, e.g., in modulo--additive channels, like the BSC).
If neither of these conditions hold (i.e., if the code is not constant
composition and the channel is not symmetric), we keep the choice $\calE_C(x,y)$
as being proportional to $-\ln W(y|x)$.

For a given choice of the Hamiltonians $\calE_S$ and $\calE_C$, in view of
these considerations, let us define the {\it joint source--channel partition
function} as the denominator of the posterior distribution, i.e.,
$$Z(\beta|\by)\dfn\sum_{\bs\in\calS^N} \exp\{-\beta[\calE_S(\bs)+
\calE_C(\bx(\bs),\by)]\}.$$
In the course of studying the properties of a  
typical realization of the joint source--channel partition function,
pertaining to a given code ensemble, 
we will make a few observations, which were already mentioned briefly
in the Introduction:
\begin{enumerate}
\item Similarly as results that have already been observed in the context
of the pure channel coding problem \cite{MM08}, 
the statistical--mechanical system pertaining to $Z(\beta|\by)$ undergoes
a phase transition, which corresponds, in the realm of
coded systems, to the transition between reliable and unreliable communication,
namely, the point at which the entropy rate of the source exceeds 
the mutual information between the input and the output of the channel.
\item When identifying the set of source 
vectors $\{\bs\}$ that dominates $Z(\beta|\by)$
(i.e., those that contribute most to $Z(\beta|\by)$) above
the phase transition temperature,
one observes a situation that parallels that of thermal equilibrium
between two physical subsystems, one corresponding to the source
and the other corresponds to the channel.
To be more specific,
if $\calE(\bs,\by)=\calE_S(\bs)+\calE_C(\bx(\bs),\by)$ is thought of
as the total `energy' shared by the source and the code/channel, then the
dominant messages $\{\bs\}$ split this total average energy between the
source and the channel components
in a way that corresponds to thermal equilibrium between the two
parallel physical subsystems.
\item The balance between the {\it thermodynamical}
entropies of the two physical subsystems that lie in equilibrium, as
described in item no.\ 2, is identified with the simple relation
between the corresponding {\it Shannon} entropies of the source,
namely, the unconditional source entropy and the conditional
entropy given the channel output, whose difference is the mutual information
between the source and the channel output. This gives rise to a simple
formula of the mutual information rate induced by a typical code in the
ensemble.
\end{enumerate}

In analogy to the definitions and the assumptions outlined in 
Section 3, we now make a few definitions and assumptions
concerning the joint source--channel coding model. 
\begin{itemize}
\item [A.1] Defining
$$\Omega_{N,\delta}^{(S)}(\epsilon)\dfn\bigg|\left\{\bs\in\calS^N:~
(\epsilon-\delta/2)N
\le \calE_S(\bs)\le 
(\epsilon+\delta/2)N\right\}\bigg|,$$
our first assumption is that
$$\Sigma_S(\epsilon)\dfn\lim_{\delta\to 0}
\lim_{N\to\infty}\frac{\ln\Omega_{N,\delta}^{(S)}(\epsilon)}{N}$$
exists and that $\Sigma_S(\epsilon)$ is a differentiable concave function.
\item [A.2] For a given $\by$, define
$$\phi_{n,\delta}(\epsilon|\by)\dfn\frac{1}{n}\ln\mbox{Pr}\{
n(\epsilon-\delta/2)\le \calE_C(\bX,\by)\le
n(\epsilon+\delta/2)\},$$
where the random vector $\bX$ is drawn under the random coding distribution
$M$, independently of $\by$. Then, our second assumption is that for all $\epsilon \ge 0$,
$\lim_{\delta\to 0}\lim_{n\to\infty}\bE\{\phi_{n,\delta}(\epsilon|\bY)\}$ tends
uniformly
to a differentiable function
$\phi(\epsilon)$, where the expectation $\bE$ is w.r.t.\ both the random
selection of the codebook and the random actions of the source and the
channel. Moreover, we assume that
$\lim_{\delta\to 0}\lim_{n\to\infty}\phi_{n,\delta}(\epsilon|\bY)$ tends
$\phi(\epsilon)$ uniformly almost surely.
\item [A.3] Let $\Sigma_S(\epsilon)$ and
$\phi(\epsilon)$ be defined as above, and let
$\Sigma_0(\epsilon)$ be defined by
the weighted supremal convolution
$$\Sigma_0(\epsilon)\dfn\max_{0\le\epsilon'\le(1+\lambda)\epsilon}
\left[\frac{\Sigma_S(\epsilon')}{1+\lambda}+\frac{\lambda}{1+\lambda}\phi\left(
\frac{(1+\lambda)\epsilon-\epsilon'}{\lambda}\right)\right].$$
Our third assumption is that $\Sigma_0(\epsilon)$ is a concave function
throughout the range of $\epsilon$ where it is non--negative. We now
define
$$\Sigma(\epsilon)=\left\{\begin{array}{ll}
\Sigma_0(\epsilon) & \Sigma_0(\epsilon)\ge 0\\
-\infty & \Sigma_0(\epsilon)< 0\end{array}\right.$$
\end{itemize}
As we shall see below,
while $\Sigma_0(\epsilon)$ gives the logarithm of the {\it expected number} of
configurations with total energy $\epsilon$, the function $\Sigma(\epsilon)$
gives the number of such configurations 
for a {\it typical} code in the ensemble. To see this,
note that if $\Sigma_S(\epsilon')+\lambda\phi([(1+\lambda)\epsilon-\epsilon']/
\lambda) < 0$ for all $\epsilon'$, this means that for every $\epsilon'$
the product of the number of configurations $\{\bs\}$ for which
$\calE_S(\bs)$ is about $n\epsilon'$ and the probability that a randomly
chosen codeword would provide the complementary energy 
$([(1+\lambda)\epsilon-\epsilon']/\lambda$, is less than one, which means that
there is a very low probability to find any configuration with total energy
$\epsilon$, and so, $\Sigma(\epsilon)$ which is the normalized logarithm of
the number of such configurations 
(i.e., the thermodynamical entropy of the combined system)
is equal to $-\infty$ for a typical code
realization. Note that the concavity of $\Sigma_0(\epsilon)$ across the 
range where it is non--negative implies that $\Sigma(\epsilon)$ is concave
as well.

In analogy to the discussion of the previous section,
let us define 
$$Z_S(\beta)\dfn\sum_{\bs}\exp\{-\beta\calE_S(\bs)\}.$$
Then,
$$\psi_S(\beta)\dfn\lim_{N\to\infty}\frac{1}{N}\ln Z_S(\beta)$$
and $\Sigma_S(\epsilon)$ are a Legendre--transform pair.
Since $\Sigma_S(\cdot)$ is assumed concave, then the inverse transform
relation
$$\Sigma_S(\epsilon)=\inf_{\beta\ge 0}[\beta\epsilon+\psi_S(\beta)],$$
holds true as well, and so the derivatives $\beta_S(\epsilon)\dfn
d\Sigma_S/d\epsilon$ and $\epsilon_S(\beta)=-d\psi_S/d\beta$
are inverses of each other. It follows then that
the Shannon entropy rate $H(S)$ of $P(\bs)$ (which depends on $\beta$)
agrees with $\Sigma_S(\epsilon)$ whenever $\beta$ and $\epsilon$ are
related by $\beta=\beta_S(\epsilon)$, or equivalently, $\epsilon=\epsilon_S
(\beta)$. 

Referring to the partition function $Z(\beta|\by)$, let us 
distinguish between the contribution of the actual realization of
the true sequence that the source actually emitted $\bs_0$, i.e., 
$Z_c(\beta|\by)=\exp\{-\beta[\calE_S(\bs_0)+\calE_C(\bx(\bs_0),\by)]\}$
and the contribution of all other (erroneous) source vectors
$$Z_e(\beta|\by)=\sum_{\bs\ne\bs_0}
\exp\{-\beta[\calE_S(\bs)+\calE_C(\bx(\bs),\by)]\}.$$
Now, $\ln Z_c(\beta|\by)$ is typically around
$-[\bE\{\calE_S(\bS)\}+\bE\{\calE_C(\bX(\bS),\bY)\}].$
As for $Z_e(\beta|\by)$, let us define 
$$\Omega_{N,\delta}(\epsilon|\by)=\bigg|\left\{\bs\ne\bs_0:~
N(1+\lambda)(\epsilon-\delta/2)\le\calE_S(\bs)+\calE_C(\bx(\bs),\by)\le
N(1+\lambda)(\epsilon+\delta/2)\right\}\bigg|.$$
Then, similarly as in the previous section,
one readily observes that for $\delta'=\delta(1+\lambda)$,
we have:
\begin{eqnarray}
\Omega_{N,\delta'}(\epsilon|\by)&=&
\sum_{j\ge 0}\Omega_{N,\delta}^{(S)}((j+1/2)\delta)\times\nonumber\\
& &\mbox{Pr}\{N(1+\lambda)(\epsilon-\delta'/2)-
N(j+1)\delta\le \calE_C(\bX,\by))\le 
N(1+\lambda)(\epsilon+\delta'/2)-Nj\delta\}\nonumber\\
&=&
\sum_{j\ge 0}\Omega_{N,\delta}^{(S)}((j+1/2)\delta)
\exp\{n\phi_{n,\delta}([(1+\lambda)\epsilon-(j+1/2)\delta]/\lambda|\by)\}
\end{eqnarray}
Taking logarithms of both sides, 
dividing by $N+n=N(1+\lambda)$, letting $N$ grow without bound,
and finally letting $\delta$ go to zero, we
obtain\footnote{At this point, we are using the fact
\cite{MM08},\cite{Merhav08} that for an ensemble of independently selected
codewords, the number of codewords which contribute energy
$\calE_C(\bX,\by)\approx
n[(1+\lambda)\epsilon-\epsilon']\lambda$, is with very high probability zero,
if $\Sigma_S(\epsilon')+\lambda\phi(1+\lambda)\epsilon-\epsilon']/\lambda) < 0$
and around
$\exp\{N[\Sigma_S(\epsilon')+\lambda\phi(1+\lambda)\epsilon-\epsilon']/\lambda)\}$
if $\Sigma_S(\epsilon')+\lambda\phi(1+\lambda)\epsilon-\epsilon']/\lambda) >
0$. The assumption of independnent codewords can be relaxed as long as this
concentration property continues to hold.} that:
$$\lim_{N\to\infty}\frac{\ln \hat{\Omega}_{N,\delta'}(\epsilon|\bY)}{N(1+
\lambda)}\eas\left\{\begin{array}{ll}
\Sigma_0(\epsilon) & \Sigma_0(\epsilon)\ge 0\\
-\infty & \Sigma_0(\epsilon) < 0\end{array}\right.$$
but the r.h.s.\ is exactly $\Sigma(\epsilon)$.
Thus, as explained earlier, 
$\Sigma(\epsilon)$ is the thermodynamical entropy associated
with the combined source--channel system.
The concavity of $\Sigma(\epsilon)$ then
implies that it agrees (after the appropriate scaling)
with the conditional Shannon entropy rate of the
source given the channel output, $H(S|Y)$, i.e., the entropy
rate pertaining to the sequence of conditional probabilities $P(\bs|\by)$
defined above. For a given $\epsilon$ in the range where $\Sigma(\epsilon)$
is finite,
let $\epsilon'=\epsilon^*$ achieve the supremum defining $\Sigma(\epsilon)$.

At this point, one should distinguish between two situations: In the first
situation, $\epsilon$ is on the boundary of the range where
$\Sigma(\epsilon )$ is finite and positive, namely, $\Sigma(\epsilon)=0$.
In this case, the partition function $Z(\beta|\by)$ 
(and hence also $P_\beta(\bs|\by)$) is dominated by a
subexponential number of configurations $\{\bs\}$ and so, the entropy rate
$H(S|Y)=0$, which means that the system is frozen in its {\it glassy phase}
(cf.\ \cite{MM08},\cite{Merhav08} and references therein.)
In the second situation, $\epsilon$ is an internal point of the range
where $\Sigma(\epsilon) > 0$, where we will also assume 
that $\epsilon^*\in(0,(1+\lambda)\epsilon)$, which is the {\it 
paramagnetic phase}
(or the disordered phase) of $Z_e(\beta|\by)$.
Then, the derivative of the function being maximized vanishes, i.e.,
$$\frac{d\Sigma_S(\epsilon')}{d\epsilon'}\Bigg|_{\epsilon'=\epsilon^*}
-\frac{d\phi(\epsilon'')}{d\epsilon''}\Bigg|_{\epsilon''=
[(1+\lambda)\epsilon-\epsilon^*]/\lambda}=0$$
or equivalently,
\begin{equation}
\label{equilibrium}
\Sigma_S'(\epsilon^*)
=\phi'\left(
\frac{(1+\lambda)\epsilon-\epsilon^*}{\lambda}\right),
\end{equation}
where $\Sigma_S'$ and $\phi'$ denote the derivatives of
$\Sigma_S$ and $\phi$, respectively. As before,
eq.\ (\ref{equilibrium}) gives rise to thermal equilbrium
between the physical system corresponding to the source
and the one that pertains to the code/channel. Next observe that
the left--hand side is exactly $\beta_S(\epsilon^*)$.
Thus,
$$\beta_S(\epsilon^*)=
\phi'\left(
\frac{(1+\lambda)\epsilon-\epsilon^*}{\lambda}\right),$$
which means that given the value of the total per--particle energy
$\epsilon$, we can find how the dominant codewords split the energy
between the source and the channel: we can solve the above equation
with the given $\epsilon$, with $\epsilon^*$ as an unknown. Then,
the source contribution will be $\epsilon^*$ and the channel contribution
will be $[(1+\lambda)\epsilon-\epsilon^*]/\lambda$. 

The discussion above holds for every value of $\epsilon$ for
which $\Sigma(\epsilon) > 0$. The dominant value of $\epsilon$
is $\epsilon_0$, the one that achieves
$\bE\{\ln Z(\beta|\bY)\}/[N(1+\lambda)]$ for large $N$, in other words,
the achiever of:
$$\psi(\beta)=\lim_{N\to\infty}\frac{\bE\ln Z(\beta|\bY)}{N(1+\lambda)}
=\sup_{\epsilon\ge 0}[\Sigma(\epsilon)-\beta\epsilon].$$
Thus, the dominant value of $\epsilon$, which is relevant for the
previous paragraph, is $\epsilon_0$, which in turn depends only on $\beta$.
But since $\Sigma$ is assumed concave, 
then $\psi$ and $\Sigma$ are also a Legendre--transform pair, and so
$\epsilon_0$ and $\beta$ are related via the derivatives,
$\epsilon_0=\epsilon(\beta)\dfn -\psi'(\beta)$ and
$\beta=\beta(\epsilon)=\Sigma'(\epsilon)$, where again, primes denote
the derivatives. In summary, given $\beta$, $\epsilon_0=\epsilon(\beta)$
and $\epsilon^*=\epsilon_S(\beta)$. Thus, $\beta_S(\epsilon^*)$
in the equilibrium equation is $\beta_s(\epsilon_S(\beta))\equiv\beta$
since $\beta_S(\cdot)$ and $\epsilon_S(\cdot)$ are inverses of one another.
Thus, the equilibrium equation applied to the dominant energy
$\epsilon_0$ becomes
$$\beta=\Sigma_S'(\epsilon^*)=
\phi'\left(
\frac{(1+\lambda)\epsilon_0-\epsilon^*}{\lambda}\right).$$
If, in addition, $\phi$ is concave, then $\phi'$ is monotone,
and thus has an inverse, which is given by the negative derivative $-\zeta'$
of the Legendre transform of $\phi$, that is,
by the derivative of 
$$\zeta(t)=\sup_{\epsilon}[
\phi(\epsilon)-\epsilon t]$$
and then 
$$\frac{(1+\lambda)\epsilon_0-\epsilon^*}{\lambda}=-\zeta'(\beta).$$

Now observe that if, for a typical
$\by$, either
$Z_c(\beta|\by)$ dominates $Z_e(\beta|\by)$, or $Z_e(\beta|\by)$ is in its
frozen phase, then $H(S|Y)$ vanishes, and so the mutual information
rate $\lim_{N\to\infty}I(\bS;\bY)/N=H(S)$. For the complementary case,
our main result is the following:
\begin{theorem}
Let $\bE\{I(\bS;\bY)\}$ denote the expected mutual
information, where the expectation is taken w.r.t.\ the ensemble of of joint
source--channel codes. Then, under Assumptions A1--A3:
$$\lim_{N\to\infty}\frac{\bE\{I(\bS;\bY)\}}{N}=-\lambda\phi(-\zeta'(\beta)),$$
provided that $\Sigma(\epsilon_0)>0$.
\end{theorem}

\noindent
{\it Remark:} From the above discussion, it is apparent that this result
applies also to the almost--sure limit of $I(\bS;\bY)/N$ w.r.t.\ the code
ensemble.

\noindent
{\it Proof.}
\begin{eqnarray}
\lim_{N\to\infty}\frac{\bE I(\bS;\bY)}{N}&=&
H(S)-H(S|Y)\nonumber\\
&=&\Sigma_S(\epsilon^*)-(1+\lambda)\Sigma(\epsilon_0)\nonumber\\
&=&-\lambda\phi\left(\frac{(1+\lambda)\epsilon_0-
\epsilon^*}{\lambda}\right)\nonumber\\
&=&-\lambda\phi(-\zeta'(\beta)). ~~~~~~~~~~~~~\Box
\end{eqnarray}

\vspace{0.25cm}

\noindent
{\it Discussion.}
We obtained then a very simple formula which depends solely on the
random coding distribution. 
But what is the meaning of $\zeta'(\beta)$?
Since $-\phi(\epsilon)$ is, in fact, 
the large deviations
rate function for the event $\calE_C(\bX,\by)\le n\epsilon$,
and $\zeta(t)$ is its Legendre transform, then
it must be the almost--sure limit of the log--moment generating function,
that is
$$\zeta(t)\eas\lim_{n\to\infty}\frac{1}{n}\ln 
\sum_{\bx\in\calX^n}M(\bx)e^{-t\calE_C(\bx,\bY)}$$
where, as defined above, $M$ is the random coding distribution
that governs each one of the independent, randomly selected codewords.
Thus,
$$-\zeta'(\beta)\eas\lim_{n\to\infty}\frac{1}{n}\cdot\frac
{\sum_{\bx}M(\bx)\calE_c(\bx,\bY)e^{-\beta\calE_C(\bx,\bY)}\}}{\sum_{\bx}
M(\bx)e^{-\beta\calE_C(\bx,\bY)}}.$$
But the Boltzmann weight
$e^{-\beta\calE_C(\bx,\by)}$ is proportional to $W(\by|\bx)$, and so,
$-\zeta'(\beta)$ is exactly the 
asymptotic almost--sure normalized conditional expectation of the energy,
$\lim_{n\to\infty}\bE\{\calE_C(\bX,\bY)|\bY\}/n$,
stemming from
the action of the channel on the message $\bx(\bs_0)$ that was actually
transmitted. This quantity in turn is assumed to concentrate about its mean
which is $\lim_{n\to\infty}\bE\{\calE_C(\bX,\bY)\}/n$.
Thus, $Z_e(\beta|\by)$ and 
$P(\bs|\by)$ are dominated by (erroneous) sequences $\{\bs\}$ whose
normalized energy $\epsilon_0$ consists of a source contribution
$\epsilon^*=
\lim_{N\to\infty}\bE\{\calE_S(\bS)\}/N$, and a channel contribution, 
$[(1+\lambda)\epsilon_0-\epsilon^*]/\lambda$ that agrees with the normalized
energy generated by the noise, i.e., it agrees
with $\lim_{n\to\infty}\bE\{\calE_C(\bX,\bY)\}/n$, where $\bX$ and $\bY$ 
are related
via the channel $W$. Moreover, this is also the typical energy composition
of the true message $\bs_0$ that was actually transmitted
(cf.\ the definition of $Z_c(\beta|\by)$. Thus, the
above conclusion holds true regardless of whether or not the 
entropy rate of the source is smaller (in which case
$\bs_0$ dominates $Z(\beta|\by)$) or larger than $\lambda$ times the
normalized mutual informtion between $\bX$ and $\bY$
(in which case, erroneous messages dominate $Z(\beta|\by)$ for a typical $\by$).
We have already seen this behavior in
the special case of the binary source and the BSC.

\vspace{0.25cm}

\noindent
{\it Example 1.} Suppose that the channel is BSC and codewords
are generated by fair coin tossing. In this case, 
$W(\by|\bx)$
is proportional to $\exp\{-\beta \calE_C(\bx,\by)\}$, where $\calE_C(\bx,\by)$
is the Hamming distance and $\beta=\ln\frac{1-p}{p}$.
In this case, $\phi(\epsilon)=h_2(\epsilon)-\ln
2$ whose derivative is $\phi'(\epsilon)=\ln\frac{1-p}{p}$, and so,
$-\zeta'(\beta)$, the inverse of $\phi'(\epsilon)$, 
is given by $-\zeta'(\beta)=1/(1+e^\beta)=p$. It follows then that
if, in addition, the source is binary and memoryless with a parameter $q$,
then $P(\bs|\by)$ is dominated by vectors $\{\bs\}$ whose energy
is as described in the Introduction.
Also, the normalized mutual
information is $-\lambda\phi(-\zeta'(\beta))=
-\lambda\phi(p)=\lambda(\ln 2-h_2(p))$. 
Somewhat more generally, let each
coordinate $X_i(\bs)$, $i=1,\ldots, n$, of each codeword be drawn i.i.d.\ with
probabilities $\mbox{Pr}\{X_i(\bs)=1\}=1-\mbox{Pr}\{X_i(\bs)=0\}=m$. Then, it
is easy to show (using the method of types \cite{CK81}) that
$$-\phi(p)=\min_{\{P_{X|Y}:~\bE d(X,Y)\le
p\}}[I(X;Y)+D(P_X\|M)],~~~~Y\sim\mbox{Bernoulli}(m\ast p),$$
where $m\ast p$ means the binary convolution between $m$ and
$p$ (i.e., $m\ast p=m(1-p)+p(1-m)$), $d(\cdot,\cdot)$
is the Hamming distance and $P_X$ is the marginal of
$X$ induced by $Y$ (which is Bernoulli($m\ast p$)) and the reversed channel $P_{X|Y}$
to be optimized. By eliminating the divergence term, we are lower bounding
$-\phi(p)$ by the rate--distortion function of $Y$ at Hamming distortion $p$,
which is $h_2(m\ast p)-h_2(p)$. On the other hand, 
returning to the original minimization problem, by selecting $P_{X|Y}$
(instead of minimizing over $P_{X|Y}$)
to be the reverse channel induced by $M$ and $W_{Y|X}$ (which is the
BSC($p$)), we are getting the same quantity also as an upper bound. Thus,
$-\phi(p)=h_2(m\ast p)-h_2(p)$, and so,
$$\lim_{N\to\infty}\frac{\bE I(\bS;\bY)}{N}=\lambda[h_2(m\ast p)-h_2(p)].$$

\vspace{0.5cm}

\noindent
{\it Comment:} An alternative view on the derivation of the
asymptotic mutual information rate between $\bS$ and $\bY$
comes from the following chain of equalities:
\begin{eqnarray}
\label{alternative}
\lim_{N\to\infty}\frac{\bE I(\bS;\bY)}{N}&=&
\lim_{N\to\infty}\bE\left\{\ln\frac{P(\bY|\bS)}{P(\bY)}\right\}\nonumber\\
&=&\lim_{N\to\infty}\frac{1}{N}
\bE\left\{\ln \exp\{-\beta\calE_C(\bX(\bS),\bY)\}\right\}-\nonumber\\
& &\lim_{N\to\infty}\frac{1}{N}\bE\left\{\ln\left[\sum_{\bs}\frac{1}{Z_S(\beta)}
\exp\{-\beta[\calE_S(\bs)+\calE_C(\bX(\bs),\bY)]\}\right]\right\}\nonumber\\
&=&-\beta[(1+\lambda)\epsilon_0-\epsilon^*]+\psi_S(\beta)
-\Sigma_S(\epsilon^*)-\lambda\phi\left(\frac{(1+\lambda)\epsilon_0-\epsilon^*)}
{\lambda}\right)+\beta(1+\lambda)\epsilon_0\nonumber\\
&=&\beta\epsilon^*+\psi_S(\beta)-\Sigma_S(\epsilon^*)-
\lambda\phi\left(\frac{(1+\lambda)\epsilon_0-\epsilon^*)}
{\lambda}\right)\nonumber\\
&=&-\lambda\phi\left(\frac{(1+\lambda)\epsilon_0-\epsilon^*)}
{\lambda}\right)
\end{eqnarray}
where we have used the fact that the summation over $\bs$ is dominated
by configurations with per--particle energy $\epsilon_0$, which is
allocated as $\epsilon^*$ and $[(1+\lambda)\epsilon_0-\epsilon^*]/\lambda$.

\section{Application to the Wiretap Channel}

In this section, we demonstrate how our results apply to the wiretap channel.
Wyner, in his well--known paper on the wiretap channel
\cite{Wyner75}, studied the problem of secure communication
across a degraded broadcast channel, without using a secret key,
where the legitimate receiver has access to the output of the good channel
and the wiretapper receives the output of the bad channel. In that paper,
Wyner characterized the optimum trade--off
between reliable coding rates and the equivocation at the
wiretapper, which was defined in terms of
the conditional entropy of the source given the
output of the bad channel, observed by the
wire--tapper. 

Consider a DMS $P$ as before, and a
cascade of two finite alphabet 
DMC's: $W_{Y|X}$ followed immediately 
by $W_{Z|Y}$, both\footnote{The notation of the output of the second channel,
$Z$, should not be confused with the notation of the partition function since
we do not refer the partition function in this section.}
operating at a relative
rate of $\lambda$ channel symbols 
per source symbol. The source $\bs\in\calS^N$
is encoded to a channel input vector $\bx(\bs)\in\calX^n$, $n=\lambda N$,
and then transmitted. A code for the wire--tap channel
should be designed in a way, that on the one hand,
the legitimate receiver
is required to estimate the source
$\bs$ from the output $\by\in\calY^n$ of the channel $W_{Y|X}$ within an
arbitrarily small probability of error, whereas 
on the other hand, the eavesdropper,
which has access to $\bz\in\calZ^n$, should be able to learn as little
as possible about the source in the sense that 
the asymptotic equivocation, 
$\Delta=\limsup_{N\to\infty}H(\bS|\bZ)/N$, should be as large
as possible. Wyner showed
\cite{Wyner75} that the largest achievable value of $\Delta$
is given by $\lambda\Gamma(H(S)/\lambda)$, where
$$\Gamma(R)\dfn\max_{P_X:~I(X;Y)\ge R}[I(X;Y)-I(X;Z)].$$
In particular, the secrecy capacity $C_s$, which is the solution to the
equation $R=\Gamma(R)$, is the rate at which the potential secrecy
that the wiretap channel can offer is fully expoilted: If the entropy of the
source, $H(S)/\lambda$ 
is less than or equal to $C_s$ (supposing that $\lambda$
can be chosen in such a way), then
the coding scheme of \cite{Wyner75} 
that asymptotically achieves $C_s$ works as follows:
Let $X^*$ be the random variable $X$ that achieves $\Gamma(R)$, for some $R$
in the range
$H(S)/\lambda \le R \le C_s$,
and let $Y^*$ and $Z^*$ be the corresponding outputs of the two channels.
We first compress the source $\bS$ to its entropy, and then apply channel
coding so that the good receiver 
can still decode reliably for large $N$ and $n$, 
but the bad one cannot. Now, since $H(S)/\lambda\le C_s$,
then by the definitions of $\Gamma(\cdot)$ and $C_s$, 
$I(X^*;Y^*)\ge H(S)/\lambda+I(X^*;Z^*)$. Accordingly, 
the channel codebook is composed of 
about $e^{NH(S)}=e^{nH(S)/\lambda}$
bins (one for each typical source sequence), each of 
size slightly less than $e^{nI(X^*;Z^*)}$. The codeword actually
transmitted is randomly chosen among all codewords of the bin pertaining
to the index of the compressed source sequence. Note that
the eavesdropper could have decoded the message had it been informed
of the bin which the transmitted codeword belongs to since the rate of
the bin, as said, is (slightly) less than $I(X^*;Z^*)$. 
The idea then is that this information
is irrelevant since it is independent of the source vector, and so
it does not help the eavesdropper in learning anything about the source.
Indeed, if we represent the transmitted codeword $\bx$ as $f(c(\bs),\bu)$,
where $c(\bs)$ stands for the bit string of the lossless compression of
$\bs$, indicating the bin index using $nH(S)/\lambda$ nats, and $\bu$ 
as an independent random bit string of length $nI(X^*;Z^*)$ nats, then
we have the following: One the one hand,
$$H(\bX|\bZ)\le H(c(\bS),\bU|\bZ)=H(c(\bS)|\bZ)+H(\bU|\bZ,c(\bS))$$
where the term $H(\bU|\bZ,c(\bS))$ essentially vanishes since, as mentioned
above, every bin forms a channel sub--code that is reliably decodable
by the eavesdropper. On the other hand,
$$H(\bX|\bZ)=H(\bX)-I(\bX;\bZ),$$
thus the equivocation achieved is:
$$H(\bS|\bZ)\ge H(c(\bS)|\bZ)\sim H(\bX)-I(\bX;\bZ)$$
where the first term in the r.h.s.\ 
is essentially $n[H(S)/\lambda+I(X^*;Z^*)]$ and the second term,
which is a mutual information induced by a code above capacity,
can be evaluated using our above results, provided that the channel code
is randomly selected
from an ensemble that satisfies our assumptions. For example, if the
codewords are chosen i.i.d.\ according to the distribution of $X^*$,
then $I(\bX;\bZ)$ is approximately $nI(X^*;Z^*)$, and then
full secrecy is achieved as $H(\bS|\bZ)/N$ is essentially equal to $H(S)$.
Nonetheless, since the rate of the code $[H(S)/\lambda+I(X^*;Z^*)$ is less
than $I(X^*;Y^*)$, the legtimate decoder can still decode reliably.
Out results can be used also to assess the secrecy achieved by random
varlaibles other than i.i.d.\ according to $X^*$, while ensuring that the good decode can still
decode reliably.

\section{Extension to Multiuser Settings}

The above ideas can be extended in a natural manner to multiuser
communication situations, and in this section, we demonstrate this
for the multiple access channel (MAC), where the underlying principle
is again thermal equilibrium between the subsystems pertaining to the
different users and that of the channel. As before, our focus is on
the regime where reliable communication cannot hold (the paramegnetic phase).

As an example, consider a randomly selected 
joint source--channel code for a MAC with 
two users, in the following setting. 
We are given two independent sources, $S_1,S_2,\ldots$ and $T_1,T_2,\ldots$ 
governed by probability distributions $P_S(\cdot)$ and $P_T(\cdot)$, which are
proportional to $\exp\{-\beta\calE_S(\cdot)\}$ and to
$\exp\{-\beta\calE_T(\cdot)\}$, 
with partition functions $Z_S(\beta)$ and $Z_T(\beta)$, respectively. Each
$N$--vector of the first source $\bs=(s_1,\ldots,s_N)\in\calS^N$ is encoded
into a channel input vector $\bx_S(\bs)\in\calX_S^n$ and each
$N$--vector of the second source $\bt=(t_1,\ldots,t_N)\in\calT^N$ is encoded
into a channel input vector $\bx_T(\bt)\in\calX_T^n$. Both codebooks are
selected independently, where each codevector of the first code is chosen
independently according to distribution 
$M_S$ and each codevector of the second codebook is
selected independently according to distribution $M_T$. Both codewords are
fed into a memoryless MAC $W(\by|\bx_S,\bx_T)$, which is proportional to
$\exp\{-\beta\calE_C(\bx_S,\bx_T,\by)\}$.
If we wish to estimate the
mutual information $\bE I(\bS,\bT;\bY)$ induced by the code, this is quite a trivial
extension of the former derivation. But what about $\bE I(\bS;\bY)$?

Here, it will be more convenient to adopt the alternative derivation
of eq.\ (\ref{alternative}). Considering the partition function
$$Z(\beta|\by)=\sum_{\bs,\bt}\exp\{-\beta[\calE_S(\bs)+\calE_T(\bt)+
\calE_C(\bx_S(\bs),\bx_T(\bt),\by)]\},$$
let $\epsilon_S^*$, $\epsilon_T^*$, and $\epsilon_C^*$ denote the
dominant energies allocated to the source $S$, the source $T$, and the
MAC, respectively. Also, for a typical randomly chosen codeword $\bx_S(\bs)$ 
of the source message $\bs$ actually transmitted,
let us
define $e^{n\phi_{n,\delta}(\epsilon|\bx_S(\bs),\by)}$ as the probability 
(under $M_T$) that
$\calE_c(\bx_S(\bs),\bX_T,\by)$ is between $n(\epsilon-\delta/2)$ and
$n(\epsilon+\delta/2)$, for given $\bx_S(\bs)$ and $\by$, and assume that
as $n\to\infty$ and then $\delta\to 0$, $\phi_{n,\delta}(\epsilon|\bx_S(\bs),
\by)$ tends uniformly almost surely to a certain function 
which will be denoted by $\phi(\epsilon|S)$. Now,
\begin{eqnarray}
\lim_{N\to\infty}\frac{\bE I(\bS;\bY)}{N}&=&
\lim_{N\to\infty}\frac{1}{N}\bE\{\ln P(\bY|\bS)\}-
\lim_{N\to\infty}\frac{1}{N}\bE\{\ln P(\bY)\}\nonumber\\
&=&\lim_{N\to\infty}\frac{1}{N}\bE\left\{\ln\left[
\frac{1}{Z_T(\beta)}\sum_{\bt}\exp\{-\beta[\calE_T(\bt)+\calE_C(\bX_S(\bS),
\bX_T(\bt),\bY)]\}\right]\right\}-\nonumber\\
& &\lim_{N\to\infty}\frac{1}{N}\bE\left\{\ln\left[
\frac{1}{Z_S(\beta)Z_T(\beta)}\sum_{\bs,\bt}\exp\{-\beta[\calE_S(\bs)+
\right.\right.\nonumber\\
& &\left.\left. \calE_T(\bt)+\calE_C(\bX_S(\bs),
\bX_T(\bt),\bY)]\}\right]\right\}\nonumber\\
&=&\psi_S(\beta)+\Sigma_T(\epsilon_T^*)+\lambda\phi(\epsilon_C^*|S)
-\beta(\epsilon_C^*+\epsilon_T^*)-\Sigma_T(\epsilon_T^*)-\nonumber\\
& &\Sigma_S(\epsilon_S^*)-\lambda\phi(\epsilon_C^*|S)+\beta(\epsilon_S^*+
\epsilon_T^*+\epsilon_C^*)\nonumber\\
&=&\lambda[\phi(\epsilon_C^*|S)-\phi(\epsilon_C^*)]
\end{eqnarray}
The last line of the above chain of equalities
can be intuitively explained as follows: The term 
$-\lambda\phi(\epsilon_C^*)$ stands for
$\lim_{N\to\infty}\bE I(\bS,\bT;\bY)/N$, because of the same reasoning as
before (if we look at the pair $(\bS,\bT)$ 
as one entity). The term $\lambda\phi(
\epsilon_C^*|S)$ corresponds to the conditional mutual information rate
$\lim_{N\to\infty}\bE I(\bT;\bY|\bS)/N$
since the true $\bS$ is given and only the random codeword of $\bT$
is selected. Thus, by the chain rule of the
mutual information, the difference
gives the mutual information rate between $\bS$ and $\bY$.

\vspace{0.5cm}

\noindent
{\it Example 2.} Consider the binary modulo--2 additive MAC, $Y=X_S\oplus X_T\oplus V$,
where all variables take on values in $\{0,1\}$, $\oplus$ denotes
addition modulo 2 (XOR), and $V$ is Bernoulli with parameter
$p=\mbox{Pr}\{V=1\}$, independent of $X_T$ and $X_S$. 
Similarly as in Example 1, let the codebooks of
the two users be generated by i.i.d.\ distributions with parameters $m_S$ and
$m_T$, respectively.
Now, as before, $\epsilon_C^*=p$
and the probability that $\bX_S\oplus\bX_T$, whose components are 
Bernoulli($m_S\ast m_T$), would fall within distance
$np$ from a typical $\by$, whose components are Bernoulli($m_S\ast m_T\ast
p$),
is exponentially $e^{n[h_2(p)-h_2(m_S\ast m_T\ast p)]}$, thus 
$\phi(p)=h_2(m_S\ast m_T\ast p)-h_2(p)$. 
On the other hand, the probability
of the same event conditioned on $\bx_S$, 
is the probability that $\bX_T$ would fall within distance
$np$ from $\by\oplus\bx_S=\bx_T\oplus\bv$ (which has Bernoulli($m_T\ast p$) components), 
and thus is of the exponential order of
$e^{n\phi(p|S)}=e^{n[h_2(p)- h_2(m_T\ast p)]}$. It follows
then that 
$$\lim_{N\to\infty}\frac{\bE I(\bS;\bY)}{N}=\lambda[h_2(m_S\ast m_T\ast
p)-h_2(m_T\ast p)].$$
In the special case where $m_T=1/2$, we get
$\lim_{N\to\infty}\frac{I(\bS;\bY)}{N}=0$ regardless of $m_S$,
in agreement with intuition, as $\bX_T$ behaves like Bernoulli(1/2) noise in
the paramagnetic regime.

\end{document}